\documentstyle[aps,preprint,psfig]{revtex}
\begin{document}
\title{The superconducting order parameter and gauge dependence}
\author{Claude de Calan and Flavio S. Nogueira}
\address{Centre de Physique Th\'eorique,  
Ecole Polytechnique, 
F-91128 Palaiseau Cedex, FRANCE}
\renewcommand{\theequation}{\thesection.\arabic{equation}}
\date{Received \today}
\tighten
\maketitle

\begin{abstract}
The gauge dependence of the renormalization group 
functions of the Ginzburg-Landau model  
is investigated. The analysis is done by means of the 
Ward-Takahashi identities. 
After defining the superconducting order parameter, 
it is shown that its exponent $\beta$ is in fact gauge 
independent. This happens because in $d=3$ the Landau gauge is the 
only gauge having a physical meaning, a property not shared by the 
four-dimensional model where any gauge choice is possible. 
The analysis is 
done in both the context of the $\epsilon$-expansion and in the 
fixed dimension approach. It is pointed out the differences that arise 
in both of these approaches concerning the gauge dependence.    
\end{abstract}
\draft

\section{Introduction} \setcounter{equation}{0}

It is often stated in the literature that the superconducting order 
parameter in a Ginzburg-Landau (GL) model has no physical meaning 
because it is a non-gauge invariant quantity. This statement is obvious if 
by order parameter we mean $<\phi(x)>$, the expectation value of the 
complex scalar field in the GL model. Since physical quantities should be 
gauge invariant, $<\phi(x)>$ cannot be considered as a physical 
quantity. It is desirable, however, to have a gauge invariant definition of 
order parameter which characterizes the superconducting phase transition. 
One such definition exists already and has been proposed by Kennedy and 
King \cite{Kennedy} in the context of the lattice superconductor. The 
order parameter proposed by them is given by 
$G=\lim_{|{\bf x}-{\bf y}|\to\infty}<\phi_{\bf x}\phi_{\bf y}^{*}>$, 
where $\phi_{\bf x}$ is a lattice field. This order parameter describes a 
phase transition in the lattice GL model. However, it is not clear that this 
phase transition coincides exactly with the normal-superconducting 
transition or, in the language of particle physics, the Higgs transition.  

In this paper we will discuss the physical meaning of the superconducting 
order parameter in the context of a continuum GL model. The aim of this 
paper is to discuss in a deeper way the questions addressed by 
one of us in a recent letter \cite{Nogueira}. It will be 
shown that a gauge invariant definition of order parameter, consistent 
with the traditional definition of the critical exponent $\beta$, is in 
fact possible. Less obvious is the gauge independence of $\beta$ in 
different schemes of renormalization. For instance, the gauge dependence 
obtained through dimensional continuation, the $\epsilon$-expansion, is 
very different from the one obtained using a fixed dimension approach. 
This approach, though less controlled, gives better values to the 
critical exponents. 
In fact, the renomalization group (RG) 
calculations based on the $\epsilon$-expansion does 
not give very good results for 
the superconducting phase transition, specially in what concerns the 
type II regime. Indeed, this regime is not readily accessible by the 
$\epsilon$-expansion which predicts a weak first order transition 
whatever the value of the Ginzburg parameter $\kappa$ \cite{HLM}. As 
shown by Dasgupta and Halperin \cite{Dasgupta} using duality arguments in a 
lattice model, the weak first order scenario does not hold in the type II 
regime where the transition should be expected to be second order. 
In order to describe correctly 
the critical behavior of superconductors using the 
$\epsilon$-expansion, it is necessary to use resummations techniques such as 
the Pad\'e-Borel resummation used by Folk and Holovatch \cite{Folk}. 

The plan of this paper is as follows. In section II we will discuss the 
RG equations for the GL model and define the order parameter. In section 
III we will use the Ward-Takahashi (WT) identities to establish the 
gauge dependence of the order parameter exactly. It will be shown that 
in a fixed dimension $d=3$ RG only the Landau gauge $a=0$ has 
a physical meaning. This result will be anticipated already in section II 
by looking the 1-loop approximation to the RG functions. However, it will 
only be proved in section III to all orders in perturbation theory. Section IV 
concludes the paper.

\section{Renormalization group in the GL model}

Let us consider the following bare action for the GL model: 

\begin{equation}
\label{action}
S=\int d^d x\left[\frac{1}{4}F_{0}^2+(D_{\mu}^0\phi_{0})^{\dag}
(D_{\mu}^0\phi_{0})+\frac{M_{0}^2}{2}A_{\mu}^0 A_{\mu}^0+m_{0}^2|\phi_{0}|^2
+\frac{u_{0}}{2}|\phi_{0}|^4\right]+S_{gf},
\end{equation}
where the zeroes denote bare quantities, $F_{0}^{2}$ is a short for 
$F_{0}^{\mu\nu}F_{0}^{\mu\nu}$ and $D_{\mu}^0=\partial_{\mu}+ie_{0}A_{\mu}^0$.
The $S_{gf}$ is the gauge fixing part and is given by

\begin{equation}
S_{gf}=\int d^d x\frac{1}{2a_{0}}(\partial_{\mu}A_{\mu}^0)^2.
\end{equation}
We have added a mass to the gauge field in order to regularize the 
infrared divergences arising from the gauge field propagator. This mass 
breaks explicitly the gauge invariance (also broken by $S_{gf}$), 
which is recovered at the critical point. Adding a mass to the gauge 
field does not affect the renormalizability of the model. This should be 
contrasted with the non-abelian case, where the only way to provide a 
mass to the gauge field without destroying 
renormalizability is through the Higgs mechanism \cite{ZJ}. 

The renormalized action is obtained from the bare action by 
rewriting it in terms of renormalized quantities. This is achieved through 
the introduction of renormalization constants. We introduce the field 
renormalization in a standard way \cite{ZJ}, 
$A_{\mu}=Z_{A}^{-1/2}A_{\mu}^0$ and $\phi=Z_{\phi}^{-1/2}\phi_{0}$. 
The renormalized action including the other renormalization constants is 
given by

\begin{eqnarray}
\label{Raction}
S&=&\int d^d x\left[\frac{Z_{A}}{4}F^2+Z_{\phi}(D_{\mu}\phi)^{\dag}
(D_{\mu}\phi)+\frac{Z_{M}M^2}{2}A^2\right.\nonumber\\
&+&\left.Z_{m}m^2|\phi|^2+\frac{Z_{u}u}{2}
|\phi|^4+\frac{Z_{a}}{2a}(\partial_{\mu}A_{\mu})^2\right].
\end{eqnarray}
The renormalization constants $Z_{M}$ and $Z_{a}$ above are in fact 
superflous. Indeed, we have that $Z_{M}=Z_{a}=1$ which means that 
the corresponding terms in the action, the gauge field mass term and 
the gauge fixing term, are not renormalized. This result is easily 
checked by studying the WT identities as follows. By adding sources to the 
corresponding fields we obtain the following WT identity:

\begin{equation}
\label{WT1}
\left\{\left(M^2-\frac{1}{a}\Delta\right)\partial_{\mu}\frac{\delta}{
\delta J_{\mu}(x)}+ie\left[J^{\dag}(x)\frac{\delta}{\delta J^{\dag}(x)}-
J(x)\frac{\delta}{\delta J(x)}\right]\right\}W(J_{\mu},J^{\dag},J)=
\partial_{\mu}J_{\mu}(x),
\end{equation}
where $W=\log Z$, $Z$ being the generating functional of  
correlation functions defined by 

\begin{equation}
Z(J_{\mu},J^{\dag},J)=\int D\phi^{\dag}D\phi DA_{\mu}
\exp{\left[-S+\int d^d x(J_{\mu}A_{\mu}+J^{\dag}\phi+\phi^{\dag}J)\right]}.
\end{equation}
When the sources are zero $Z$ gives the partition function. 
$W$ is the generating functional of the connected 
correlation functions. The Legendre transform of $\Gamma$ of $W$  
is given by 

\begin{equation}
W(J_{\mu},J^{\dag},J)+\Gamma(\varphi^{\dag},\varphi,a_{\mu})=
\int d^d x(J_{\mu}a_{\mu}+J^{\dag}\varphi+\varphi^{\dag}J)
\end{equation} 
where the functional effective action 
$\Gamma(\varphi^{\dag},\varphi,a_{\mu})$  
is the generator of the 1-particle irreducible functions. It satisfies a 
WT identity which is the Legendre transform of (\ref{WT1}):

\begin{equation}
\label{WT2}
\left(\frac{1}{a}\Delta-M^2\right)
\partial_{\mu}a_{\mu}(x)+\partial_{\mu}\frac{\delta\Gamma}{\delta a_{\mu}(x)}+ie
\left[\varphi(x)\frac{\delta\Gamma}{\delta\varphi(x)}-\varphi^{\dag}(x)
\frac{\delta\Gamma}{\delta\varphi^{\dag}(x)}\right]=0.
\end{equation}
Both WT identities (\ref{WT1}) and (\ref{WT2}) are linear. Thus, if we 
perform a loop expansion, $\Gamma=\sum_{l}\Gamma_{l}$, where $\Gamma_{l}$ 
is the $l$-loop correction to the effective action, it follows that 
each $\Gamma_{l}$ satisfy separately a WT identity (\ref{WT2}). The 
zero-loop contribution $\Gamma_{0}$ is the classical action $S$, which 
satisfies the inhomogeneous WT identity (\ref{WT2}), that is, the 
identity containing the term 
$(a^{-1}\Delta-M^2)\partial_{\mu}a_{\mu}(x)$. The 
loop corrections, on the other hand, satisfy the homogeneous 
WT identity. The non-gauge invariant terms are all included in the 
inhomogeous WT identity, do not receive any correction due to the 
fluctuations and are in this way not renormalized, that is, the 
coresponding counterterms are zero. The only non-gauge invariant terms in 
the action are the gauge field mass term and the gauge fixing term. As a 
consequence of these reasonings we obtain that $Z_{M}=Z_{a}=1$. This 
implies that $M^2=Z_{A}M_{0}^2$ and $a=Z_{A}^{-1}a_{0}$. From 
(\ref{Raction}) we deduce also $m^2=Z_{m}^{-1}Z_{\phi}m_{0}^2$, 
$e^2=Z_{A}e_{0}^2$ and $u=Z_{\phi}^2 Z_{u}^{-1}u_{0}$. 

It is useful at 
this point to fix the renormalization conditions necessary 
to define the corresponding renormalized quantities:

\begin{equation}
\label{RCm}
\Gamma^{(2)}_{ij}(0)=m^2\delta_{ij},
\end{equation}

\begin{equation}
\left.\frac{\partial\Gamma^{(2)}_{ij}}
{\partial p^2}\right|_{p^2=0}=\delta_{ij},
\end{equation}

\begin{equation}
\Gamma^{(4)}_{ijkl}(0,0,0,0)=(\delta_{ij}\delta_{kl}+\delta_{ik}\delta_{jl}
+\delta_{il}\delta_{jk})u,
\end{equation}

\begin{equation}
\Gamma_{\mu\nu}(0)=M^2\delta_{\mu\nu},
\end{equation}

\begin{equation}
\left.\frac{\partial}{\partial p^2}
\left(\delta_{\mu\nu}-\frac{p_{\mu}p_{\nu}}{p^2}\right)
\Gamma^{(2)}_{\mu\nu}\right|_{p^2=0}=d-1,
\end{equation}

\begin{equation}
\left.\frac{\partial}{\partial p^2}\frac{p_{\mu}p_{\nu}}{p^2}
\Gamma^{(2)}_{\mu\nu}\right|_{p^2=0}=\frac{1}{a}, 
\end{equation}
where the latin indices represent the components of the scalar field.    

We are interested in the behavior when $m\to 0$, the infrared behavior. 
Thus, we assume that $m_{0}^2\propto t$, $t=(T-T_{c})/T_{c}$ 
being the reduced temperature, and 
$m=\xi^{-1}\sim t^{\nu}$ as $t\to 0$, $\xi$ being the correlation lenght. 
The RG equations will be obtained by differentiating with respect to 
$\ln m$ with all bare quantities fixed, except for $m_{0}^2$. Let us  
define the dimensionless couplings $f=m^{d-4}e^2$,  
$\hat{u}=m^{d-4}u$ and $v=m/M$. The following {\it exact} flow equations 
are readily obtained:

\begin{eqnarray}
\label{M}
m\frac{\partial M^2}{\partial m}&=&\eta_{A}M^2,\\
\label{a}
m\frac{\partial a}{\partial m}&=&-\eta_{A}a,\\
\label{f}
m\frac{\partial f}{\partial m}&=&(\eta_{A}+d-4)f,\\
\label{v}
m\frac{\partial v}{\partial m}&=&\left(1-\frac{\eta_{A}}{2}\right)v,
\end{eqnarray}
where we have introduced the RG function $\eta_{A}$ which 
at the fixed point gives the anomalous 
dimension of the gauge field. It is defined by

\begin{equation}
\eta_{A}=m\frac{\partial}{\partial m}\log Z_{A}.
\end{equation}
At 1-loop order and in fixed dimension $d=3$ we obtain

\begin{equation}
\eta_{A}=\frac{f}{24\pi}.
\end{equation}
From Eq. (\ref{f}) we obtain that a charged fixed point should correspond 
to $\eta_{A}=4-d$. In $d=4$ this corresponds necessarily to zero charge 
while in $d=3$ we have $f_{*}=24\pi$, a non-zero charge. The only 
fixed point to Eq. (\ref{a}) corresponding to a charged fixed point 
in $d=3$ is $a_{*}=0$, that is, the Landau gauge. Due to the 
negative sign in (\ref{a}), any $a\neq 0$ in the flow diagram will 
flow to infinity. Fig. 1 shows the flow diagram in the 
$fa$-plane in the 1-loop approximation in $d=3$. Note that the 
line $a=0$ contains an attractive flow towards the fixed point 
$f_{*}=24\pi$. Any $a\neq 0$, no matter how small, will flow away. 
This does not happen at $d=4$ where $f_{*}=0$ and $a=a_{*}$ 
arbitrary is a line of fixed points. This 1-loop calculation suggests 
that in $d=3$ the Landau gauge is the only physical gauge. 

The 1-loop beta function for the coupling $\hat{u}$ is given by

\begin{equation}
\label{betau}
\beta_{\hat{u}}=m\frac{\partial\hat{u}}{\partial m}=
(2\eta_{\phi}-1)\hat{u}+\frac{5}{8\pi}
\hat{u}^2+\frac{v}{2\pi}f^2.
\end{equation}  
This beta function does not exhibit any gauge dependence, which is 
cancelled out between the $\eta_{\phi}\hat{u}$ and $f^2$ term. In 
(\ref{betau}) $\eta_{\phi}$ is defined as 
$m\partial\ln Z_{\phi}/\partial m$, which computed in the Landau gauge 
gives

\begin{equation}
\label{etaphi}
\eta_{\phi}=
-\frac{2}{3\pi}\frac{f v^2}{(1+v)^2}.
\end{equation}
We will see in section III that the gauge independence of 
$\beta_{\hat{u}}$ is valid to 
all orders, though $Z_{\phi}$ does depend on the gauge. 

The flow equations we have obtained are similar to the ones obtained 
by Herbut \cite{HerbutI} in the context of the continuum dual GL model
\cite{Kovner,Kiometzis,KiometzisI,Tesanovic,deCN}. The 
difference is that in our case the charge $e$ is not meant 
to be the dual charge and, as a consequence, our mass $M$ has nothing 
to do with the photon mass (anyway, this is a subtle point, even 
in the dual approach; see ref. \cite{deCN}).  

Let us discuss the critical behavior that arises from the so defined 
GL model. From an experimental point of view, presently only the 
exponents $\alpha$ (specific heat) and $\nu$ (correlation lenght) are 
accessible. The order parameter exponent $\beta$ seems not to be 
directly accessible. This is the main point we would like to discuss. 
At this point, an important remark is in order. The critical exponents 
are obtained from the singular behavior of the effective action 
(the free energy) and correlation functions. The WT identities imply that 
the singular behavior is the same, whatever $t<0$ (the broken symmetry 
regime or ordered phase) or $t>0$ (symmetric regime or disordered 
phase). This means that the exponents have the same value below or 
above $T_{c}$. This remark is very important concerning the superconductors. 
The point is that below the transition the photon becomes massive, 
phenomenon known in particle physics as the Higgs mechanism \cite{ZJ}. 
This mass is generated spontaneously and not added by hand as our mass $M$. 
Since the singular behavior is the same as in the symmetric phase, 
the renormalization proceeds exactly as in the symmetric phase. Thus, this 
mechanism is very important in the context of non-abelian gauge theories, 
where adding a mass by hand destroys renormalizability. The photon mass 
generated by the Higgs mechanism in the GL model corresponds to the 
inverse of the penetration depth $\lambda$. This lenght is known to scales 
as the correlation lenght $\xi$ \cite{HerbutI,Herbut,deCN}, that is, 
$\lambda\sim\xi\sim |t|^{\nu}$. $\lambda$ is a quantity that arises only 
for $t<0$ but its exponent, being the same appearing in the scaling of 
$\xi$, can be evaluated for $t>0$ or even for $t=0$ (the critical point), 
this last case needing renormalization conditions different from the ones 
we use here due to the infrared divergences \cite{Herbut,deCalan}. 

The $\nu$ exponent is obtained easily by considering the flow of 
$m_{0}^2$ (remember that $m_{0}$ is not kept fixed under RG). From the 
definition of its renormalized counterpart $m^2=Z_{m}^{-1}Z_{\phi}m_{0}^2$,  
we obtain

\begin{equation}
\label{m0}
m\frac{\partial m_{0}^2}{\partial m}=(2+\eta_{m}-\eta_{\phi})m_{0}^2,
\end{equation}         
where

\begin{equation}
\eta_{m}=m\frac{\partial\ln Z_{m}}{\partial m}.
\end{equation}
Eq. (\ref{m0}) implies that near the phase transition ($m\to 0$) 
$m_{0}^2\sim m^{2+\eta_{m}^{*}-\eta}$, where $\eta_{m}^{*}$ and 
$\eta$ are the fixed point values of $\eta_{m}$ and $\eta_{\phi}$, 
respectively. Since by definition $m_{0}^2\propto t$ and 
$m\sim t^{\nu}$ we obtain immediately

\begin{equation}
\label{nu}
\nu=\frac{1}{2+\eta_{m}^{*}-\eta}.
\end{equation}

The $\nu$ exponent can be accurately measured through a direct measurement 
of the penetration depth \cite{Kamal}. In the accessible critical 
region, the fluctuations are governed by the 3D XY fixed point
\cite{Fisher}, which is a neutral fixed point. Then, it results that 
$\rho_{s}\sim\lambda^{-2}\sim t^{2\nu'}$, 
where $\rho_{s}$ is the superfluid density \cite{deCN,Fisher} and 
we have defined the exponent $\nu'$ of the penetration depth. 
Using the Josephson relation $\rho_{s}\sim t^{\nu(d-2)}$ 
\cite{Josephson} for the $d=3$, we obtain $\nu'=\nu/2$. The exponent 
$\nu'$ obtained experimentally in bulk samples of  
$YBa_{2}Cu_{3}O_{7-\delta}$ is $\nu'\approx 1/3$ \cite{Kamal} and 
therefore we obtain $\nu\approx 2/3$, corresponding in this way to a 
3D XY universality class. This very same value of $\nu$ is expected for the 
charged transition, the so called ``inverted'' 3D XY universality class 
\cite{Dasgupta}. In fact, this is confirmed by recent RG calculations 
\cite{Nogueira,Herbut,HerbutI,Kiometzis,deCalan} and Montecarlo 
simulations \cite{Olsson}. Let us evaluate $\nu$ from Eq. (\ref{nu}) in 
the present framework. We have at 1-loop order that 

\begin{equation}
\eta_{m}=-\frac{f}{4\pi}.
\end{equation}
The infrared stable fixed point corresponds to $f_{*}=24\pi$, 
$\hat{u}_{*}=8\pi/5$ and $v_{*}=0$. This gives $\eta=0$  
and $\nu=0.63$ at 1-loop order. This result can be systematically 
improved. Indeed, the fixed point $\hat{u}_{*}$ will always have a 
3D XY value because every power of $f$ will be multiplied by a function of 
$v$ which goes to zero as $m\to 0$, since $v\sim m^{1/2}$. Therefore, 
we obtain already at 2-loops $\nu\approx 2/3$. It should be noted that 
this behavior is obtained because from Eq. (\ref{M}) we have that 
$M^2\sim m^{4-d}$ near the phase transition. For $d=3$ this means 
$M\sim t^{\nu/2}$. Note that this is exactly the scaling behavior of 
the true photon mass near the neutral 3D XY fixed point! This means 
that $M$ can be alternatively 
regarded as a photon mass describing the crossover 
regime observed experimentally. This is the 
case if we interpret the GL model given in 
(\ref{action}) as a disorder field theory, that is, as a  
continuum dual GL model 
\cite{HerbutI,Kovner,Kiometzis,KiometzisI,Tesanovic,deCN}. In this case, 
the charge $e$ should be replaced by $2\pi M/q$, where $q$ is a the 
dual charge satisfying the Dirac condition $qe=2\pi$. 

Let us investigate now how the exponent $\beta$ should be defined. This 
presupposes a definition of order parameter whose scaling has exponent 
$\beta$. This order parameter arises naturally from the definition of 
the superfluid density $\rho_{s}$. This is defined for $t<0$ by 

\begin{equation}
\label{rho}
\rho_{s}=<|\phi|^2>=Z_{\phi}^{-1}<|\phi_{0}|^2>.
\end{equation}
We define the order parameter by $\Phi$ by 

\begin{equation}
\label{ordpar}
\Phi=\sqrt{<|\phi_{0}|^2>}.
\end{equation}  
The expectation value of a gauge invariant operator is gauge independent 
\cite{ZJ,Collins}. 
Thus, $\rho_{s}$ should be independent of $a$. 
However, $Z_{\phi}$ is gauge dependent (see section III) and  
we have that $\Phi$ should in fact depends on $a$ to 
cancels the gauge dependence coming from $Z_{\phi}^{-1}$ in 
(\ref{rho}), otherwise $\rho_{s}$ would be gauge dependent. 
Note that the gauge invariance of $|\phi_{0}|^2$ implies that $\Phi$ 
is independent of $a_{0}$ but {\it not} necessarilly independent of 
$a$. 
However, its scaling behavior 
near the phase transition will be shown (section III) to be gauge 
independent or, more precisely, to be evaluated in the Landau gauge.  
The same will shown to be true for $Z_{\phi}$. 
For the moment, let us check that the above definition of order 
parameter works. The exponent $\beta$ is defined through 
its behavior near the critical point, $\Phi\sim|t|^{\beta}$. Also, we 
have $Z_{\phi}\sim m^{\eta}\sim|t|^{\nu\eta}$. Putting all of this 
together we obtain $\rho_{s}\sim|t|^{2\beta-\nu\eta}$, which is the 
Josephson relation as obtained originally in Josephson's 
paper \cite{Josephson}. Using the hyperscaling relation 
$d\nu=2-\alpha$ together with the 
combination of the scaling 
relations $\alpha+2\beta+\gamma=2$ and $\gamma=\nu(2-\eta)$ to eliminate 
$\gamma$, we obtain the Josephson relation in the form 
$\rho_{s}\sim|t|^{\nu(d-2)}$. This last form of the Josephson relation has 
been also 
obtained directly by us \cite{deCN}, 
without using the scaling relations 
$\alpha+2\beta+\gamma=2$ and $\gamma=\nu(2-\eta)$. 
Therefore,  

\begin{equation}
\label{beta}
\beta=\frac{\nu}{2}(d-2+\eta).
\end{equation}
Thus, a gauge dependence in $\eta$ would imply a gauge dependence 
in $\beta$ if $\nu$ is gauge independent. Note that the gauge 
independence of $\rho_{s}$ does not ensure the gauge independence of $\nu$. 
The point is that the gauge independence of $|t|^{\nu(d-2)}$ can be 
a result of compensating gauge dependences arising from $|t|$ 
(equivalently, $m_{0}^2$) and $\nu$. 

\section{The gauge dependence} \setcounter{equation}{0}

In the preceding section we discussed the critical behavior of the GL 
model while trying to get some insight on the effect of the gauge 
dependence. Clearly a more careful analysis is needed. In this section  
gauge dependence always means a dependence on $a$, the renormalized 
gauge fixing parameter. Therefore, the other renormalized parameters of the 
model are trivially gauge independent. This statement implies the 
gauge independence of all beta functions. Note that this does not means 
necessarilly that the bare parameters are gauge independent. 

In order to obtain the gauge dependence of $Z_{\phi}$ we will employ 
the WT identities. 
From the WT identity Eq. (\ref{WT1}) we obtain the following identity:

\begin{equation}
\label{eqaux1}
\left(M^2-\frac{1}{a}\partial^2_{z}\right)\partial_{\mu}^{z}
W_{\mu}^{(2)}(z;y,x)=ie\left[\delta(y-z)W^{(2)}(z,x)-\delta(z-x)W^{(2)}(
y,z)\right].
\end{equation}
By using twice Eq. (\ref{eqaux1}) we obtain

\begin{equation}
\label{WT3}
W^{(2)}_{(\partial_{\mu}A_{\mu})^2}(p)=2e^2\int\frac{d^d k}{(2\pi)^3}
\frac{a^2}{(k^2+a M^2)^2}\left[W^{(2)}(p+k)-W^{(2)}(p)\right], 
\end{equation}
where $W^{(2)}_{(\partial_{\mu}A_{\mu})^2}(p)$ is the Fourier transform 
of 

\begin{eqnarray}
W^{(2)}_{(\partial_{\mu}A_{\mu})^2}(x,y)&=&\int d^d z\left[<
(\partial_{\mu}A_{\mu})^2(z)\phi(x)\phi^{\dag}(y)>\right.\nonumber\\
&-&\left.<(\partial_{\mu}A_{\mu})^2(z)><\phi(x)\phi^{\dag}(y)>\right].
\end{eqnarray}
Let us denote the bare counterpart of $W^{(2)}_{(\partial_{\mu}A_{\mu})^2}$ 
by $W^{(2)}_{(\partial_{\mu}A_{\mu}^0)^2,0}$. We have that 

\begin{equation}
\label{eqaux2}
2a_{0}^2\frac{\partial W_{0}^{(2)}}{\partial a_{0}}(x,y)
=W^{(2)}_{(\partial_{\mu}A_{\mu}^0)^2,0}(x,y), 
\end{equation}
where $W^{(2)}_{0}(x,y)=<\phi_{0}(x)\phi_{0}^{\dag}(y)>$ is the bare 
2-point connected correlation function. Eq. (\ref{WT3}) is valid also 
if we replace the renormalized correlation functions by the bare ones and 
the renormalized couplings by their bare counterparts. Using then a bare 
version of (\ref{WT3}) and Eq. (\ref{eqaux2}), we obtain 

\begin{equation}
\label{eqcentral}
\frac{\partial W^{(2)}_{0}}{\partial a_{0}}(p)=
e^2_{0}\int\frac{d^3k}{(2\pi)^3}\frac{W^{(2)}_{0}(p+k)-W^{(2)}_{0}(p)}{
(k^2+a_{0}M^2_{0})^2}.
\end{equation}
Eq. (\ref{eqcentral}) can be rewritten as

\begin{equation}
\label{eqcentral1}
\frac{\partial\ln Z_{\phi}}{\partial a_{0}}W^{(2)}(p)+
\frac{\partial W^{(2)}}{\partial a_{0}}(p)= 
e^2_{0}\int\frac{d^3k}{(2\pi)^3}\frac{W^{(2)}(p+k)-W^{(2)}(p)}{
(k^2+a_{0}M^2_{0})^2},
\end{equation}
out of which we obtain 

\begin{eqnarray}
\label{gdepZ}
\frac{\partial\ln Z_{\phi}}{\partial a_{0}}&=&
-e^2_{0}\int\frac{d^d k}{(2\pi)^d}\frac{1}{(k^2+a_{0}M^2_{0})^2}\nonumber\\
&=&e^2_{0}a^{\frac{d-4}{2}}_{0}
M^{d-4}_{0}\left(\frac{d}{2}-1\right)C_{d},
\end{eqnarray}
where $\pi/C_{d}=(4\pi)^{d/2}\Gamma(d/2)\sin(\pi d/2)$. Note that we 
have a pole for $d=4$ in the second line of Eq. (\ref{gdepZ}). This is 
a consequence of the logarithmic divergence for $d=4$. In the 
$\epsilon$-expansion the singular part of the different correlation 
functions is isolated as poles in $1/\epsilon$ with $\epsilon=4-d$ and 
the renormalization constants are written as power series in 
$1/\epsilon$. This way of doing the things leads to the determination 
of the critical exponents as power series in $\epsilon$ \cite{ZJ}. 
The physical case of interest in critical phenomena of superfluid and 
magnetic systems corresponds to $\epsilon=1$. 

By integrating the first line of (\ref{gdepZ}) we obtain

\begin{equation}
\label{gdepZ1}
\ln Z_{\phi}(a_{0})=\ln Z_{\phi}(a_{0}=0)-e^2_{0}a_{0}
\int\frac{d^d k}{(2\pi)^d}\frac{1}{k^2(k^2+a_{0}M^2_{0})^2}.
\end{equation}
Since $e^2_{0}a_{0}=e^2a$ and $a_{0}M^2_{0}=a M^2$, 
we can rewrite Eq. (\ref{gdepZ1}) as 

\begin{equation}
\label{gdepZ2}
\ln Z_{\phi}(a)=\ln Z_{\phi}(a=0)-e^2a
\int\frac{d^d k}{(2\pi)^d}\frac{1}{k^2(k^2+a M^2)^2}.
\end{equation}
Let us assume that $Z_{\phi}(a=0)$ has been evaluated as a power 
series in $1/\epsilon$. After regularizing dimensionally the integral 
in (\ref{gdepZ2}), we obtain

\begin{equation}
\label{gdepeta}
\eta_{\phi}(a)=\eta_{\phi}(a=0)-\frac{a f}{2\pi}, 
\end{equation}
which gives the gauge dependence of $\eta_{\phi}$ in the framework of 
the $\epsilon$-expansion. Let us assume that an infrared stable 
fixed point has been obtained, for instance, by ressummation 
methods \cite{Folk}. As $m\to 0$, $f\to f_{*}\neq 0$ (if $\epsilon=1$), 
but $a$ scales as $m^{-1}$ near the fixed point and 
any non-zero $a$ runs away as $m\to 0$. Thus, the only safe way 
towards the charged fixed point is over the line $a=0$, that is, 
the Landau gauge. Note that for the case of interest in particle 
physics, $d=4$, we obtain the same equation as (\ref{gdepeta}). However, 
for $d=4$ any gauge choice is possible since $f_{*}=0$ in this 
case.

In the fixed dimension approach things work differently. For $d=3$ the 
integral in Eq. (\ref{gdepZ2}) is convergent and we can interchange 
the differentiation with respect to m with the integral sign. 
Since $a e^2$ and $a M^2$ are both RG invariants, we obtain 

\begin{equation}
\label{gdepeta1}
\eta_{\phi}(a)=\eta_{\phi}(a=0),
\end{equation}
and we obtain again that the physical gauge corresponds to $a=0$. At 
this point some remarks are in order. First, from Eq. (\ref{gdepeta1}) 
we obtain $\partial\eta_{\phi}/\partial a=0$ while the same is not 
true for the $\eta_{\phi}(a)$ given in Eq. (\ref{gdepeta}). Second, 
Eq. (\ref{gdepeta1}) can be easily checked at 1-loop order. The 
renormalization constant is given as a function of $f$, 
$a$ and $v$ and if we take care of differentiating $a$ when 
obtaining $\eta_{\phi}$, the result (\ref{gdepeta1}) follows and  
coincides with Eq. (\ref{etaphi}). Concerning the 1-loop example, it is 
instructive to ask ourselves what happens in other fixed dimension 
approaches. For instance, we could perform a critical point ($m=0$) 
calculation where the renormalization conditions are defined at  
non-zero external momenta, taking the symmetrical point for functions 
which depend on more than one momentum variable 
\cite{Herbut,deCalan}. In this case the photon mass $M$ is 
unecessary since the non-zero external momenta take care of infrared 
divergences \cite{Note}. The 1-loop expression for $Z_{\phi}$ in an arbitrary 
gauge is in this case rather simple and has been calculated by 
Schakel \cite{Schakel}. It turns out in this case 
that $Z_{\phi}$ is independent of $a$ if $d=3$. 

The gauge dependence of $Z_{m}$ can be obtained in an analogous way. 
From the renormalization condition (\ref{RCm}) and 
$m^2=Z_{\phi}Z_{m}^{-1}m_{0}^2$, we obtain that 
$W^{(2)}_{0}(0)=Z_{m}/m_{0}^2$. Using again the bare version of 
(\ref{WT3}), we obtain exactly the same equation as Eq. (\ref{gdepZ}) but 
with $Z_{\phi}$ replaced by $Z_{m}$. This means that the gauge 
dependence of $Z_{m}$ is the same as for $Z_{\phi}$. If we use 
the $\epsilon$-expansion we have that $\eta_{m}-\eta_{\phi}$ is 
gauge independent since the gauge dependence of $\eta_{m}$ will cancel 
exactly the gauge dependence of $\eta_{\phi}$. In fixed dimension 
$d=3$, on the other hand, $\eta_{m}(a)=\eta_{m}(a=0)$. It 
follows that the critical exponent $\nu$ is gauge independent. Since 
$\eta$ is gauge independent, it follows that $\beta$ is gauge 
independent and the order parameter $\Phi$ has a true physical meaning.

\section{Conclusions}

In this paper we have shown that the critical exponent $\beta$ of the 
superconducting order parameter is gauge independent or, stating 
more correctly, that it must be evaluated in the Landau gauge 
$a=0$. We may wonder if it is not a wasting of time to prove a 
result about something unaccessible experimentally. The point is that it 
is not sure that $\Phi$ cannot be measured and we hope that the discussion 
in this paper could stimulate some experimental effort in this sense. 
Moreover, we have shown that transversality (the Landau gauge) is an 
intrinsic physical feature of the $d=3$ GL model, a property not shared 
by the $d=4$ model. In $d=4$ (the case of interest in particle physics) 
the Landau gauge is used due to its computational simplicity \cite{Coleman}. 
In contrast, the Landau gauge is the only physically meaningful gauge 
in $d=3$ \cite{Note1}.   


\newpage

\begin{figure}
\centerline{\psfig{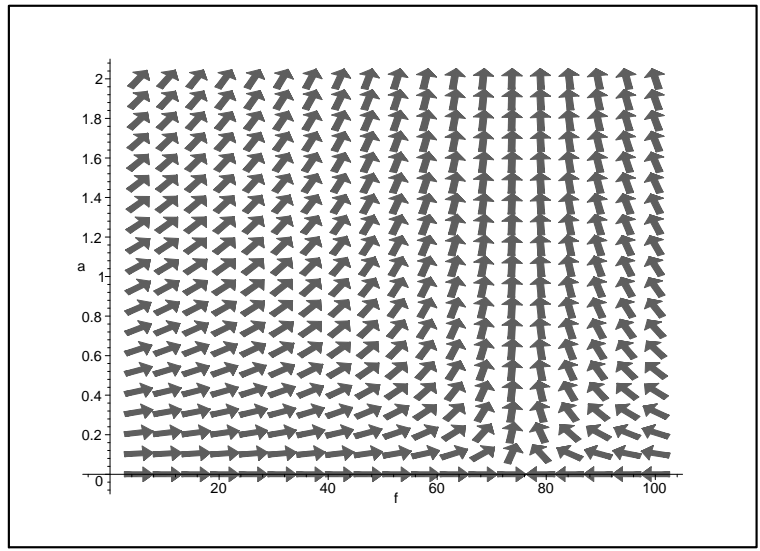}}
\caption{Flow diagram in the $fa$-plane.}
\end{figure}


\begin{references}

\bibitem{Kennedy} T. Kennedy and C. King, Phys. Rev. Lett. {\bf 55}, 776 
(1985); Commun. Math. Phys. {\bf 104}, 327 (1986); C. Borgs and 
F. Nill, {\it ibidem}, page 349.

\bibitem{Nogueira} F. S. Nogueira, Europhys. Lett. {\bf 45}, 612 (1999).

\bibitem{HLM} B. I. Halperin, T. C. Lubensky and S.-K. Ma, 
Phys. Rev. Lett. {\bf 32}, 292 (1974); J.-H. Chen, T. C. Lubensky 
and D. R. Nelson, Phys. Rev. B {\bf 17}, 4274 (1978).

\bibitem{Dasgupta} C. Dasgupta and B. I. Halperin,  
Phys. Rev. Lett. {\bf 47}, 1556 (1981).

\bibitem{Folk} R. Folk and Y. Holovatch, J. Phys. A {\bf 29}, 3409 (1996).

\bibitem{ZJ} J. Zinn-Justin, Quantum Field Theory and Critical Phenomena, 
2nd edition (Oxford, 1993).

\bibitem{HerbutI} I. F. Herbut, J. Phys. A {\bf 30}, 423 (1997).

\bibitem{Kovner} A. Kovner, P. Kurzepa and B. Rosenstein, 
Mod. Phys. Lett. A {\bf 8}, 1343 (1993).

\bibitem{Kiometzis} M. Kiometzis, H. Kleinert and A. M. J. 
Schakel, Phys. Rev. Lett. {\bf 73}, 1975 (1994).

\bibitem{KiometzisI} M. Kiometzis, H. Kleinert and A. M. J. Schakel, 
Fortschr. Phys. {\bf 43}, 697 (1995) and references therein.

\bibitem{Tesanovic} Z. Tesanovic, Phys. Rev. B {\bf 59}, 6449 (1999).

\bibitem{deCN} C. de Calan and F. S. Nogueira, Phys. Rev. B (to be published).

\bibitem{Herbut} I. F. Herbut and Z. Tesanovi\'c,  
Phys. Rev. Lett. {\bf 76}, 4588 (1996); 
I. D. Lawrie, Phys. Rev. Lett., {\bf 78}, 979 (1997); 
I. F. Herbut and Z. Tesanovi\'c, Phys. Rev. Lett. {\bf 78}, 980 (1997).

\bibitem{deCalan} C. de Calan, A. P. C. Malbouisson, F. S. Nogueira and 
N. F. Svaiter, Phys. Rev. B {\bf 59}, 554 (1999).

\bibitem{Kamal} S. Kamal, D. A. Bonn, 
N. Goldenfeld, P. J. Hirschfeld, R. Liang and 
W. N. Hardy, Phys. Rev. Lett. {\bf 73}, 1845 (1994);
S. Kamal, R. Liang, A. Hosseini, D. A. Bonn and 
W. N. Hardy, Phys. Rev. B {\bf 58}, R8933 (1998).

\bibitem{Fisher} D. S. Fisher, M. P. A. Fisher and D. A. Huse, 
Phys. Rev. B {\bf 43}, 130 (1991).

\bibitem{Olsson} P. Olsson and S. Teitel, Phys. Rev. Lett. {\bf 80}, 
1964 (1998).

\bibitem{Josephson} B. D. Josephson, Phys. Lett. {\bf 21}, 608 (1966).

\bibitem{Note} It is unecessary for computational purposes; however, 
an analysis of the gauge dependence to all orders still need the mass 
$M$ in order to avoid infrared divergences in the WT identities.

\bibitem{Collins} J. C. Collins, Renormalization (Cambridge, 1984).

\bibitem{Schakel} A. M. J. Schakel, cond-mat/9805152 (unpublished).  

\bibitem{Coleman} S. Coleman and E. Weinberg, Phys. Rev. D {\bf 7}, 
1888 (1973).

\bibitem{Note1} For $d=4$ the gauge independence 
of the S-matrix elements and particle mass ratios 
were established to all orders for the massless scalar QED 
\cite{Iliopoulos}. 
This analysis has been done in the theoretical 
framework of radiatively induced symmetry breaking \cite{Coleman} which 
describes a first order order phase transition. This type of 
phase transition can be easily obtained in $d=3$ by means of a 
gauge field fluctuation-corrected mean-field theory \cite{HLM} and is 
consistent with the weak first-order transition scenario. The same 
mean-field calculation applies in $d=4$ 
\cite{Malbouisson}. 
and reproduces the Coleman-Weinberg effective potential \cite{Coleman}. 
Unfortunately this phase transition scenario is incompatible with a 
type II regime of superconductors and the analysis of the gauge 
dependence following from it is not appropriate in describing 
thermal fluctuations associated to a second order phase transition. 

\bibitem{Iliopoulos} J. Iliopoulos and N. Papanicolau, 
Nucl. Phys. B {\bf 105}, 77 (1976).

\bibitem{Malbouisson} A. P. C. Malbouisson, F. S. Nogueira and N. F. Svaiter, 
Mod. Phys. Lett. B {\bf 11}, 749 (1996). 

\end{references}
\end{document}